\documentclass[preprintnumbers,amsmath,amssymb,floatfix,10pt,prd,onecolumn,
superscriptaddress,nofootinbib]{revtex4}
\usepackage{amsfonts}
\usepackage{mathrsfs}
\usepackage{latexsym}
\usepackage{epsfig}
\usepackage{epstopdf}
\usepackage{graphicx}
\usepackage{amssymb}
\usepackage{amsmath}
\usepackage{dcolumn}
\usepackage{bm}
\usepackage{color}
\usepackage{xcolor}
\usepackage{comment}
\usepackage[cp1251]{inputenc}
\begin{document}

\title{\bf Massive Compact Bardeen Stars with Conformal Motion}

\author{M. Farasat Shamir}
\email{farasat.shamir@nu.edu.pk; farasat.shamir@gmail.com}\affiliation{National University of Computer and
Emerging Sciences,\\ Lahore Campus, Pakistan.}

\begin{abstract}

The main focus of this paper is to discuss the solutions of Einstein-Maxwell's field equations for compact stars study. We have chosen the MIT bag model equation of state for the pressure-energy density relationship  and conformal Killing vectors are used to investigate the appropriate forms for metric coefficients.
We impose the boundary conditions, by choosing the Bardeen model to describe as an exterior spacetime. The Bardeen model may provide the analysis with some interesting results. For example, the extra terms involved in the asymptotic representations as compared to the usual Reissner-Nordstrom case may influence the mass of a stellar structure.  Both energy density and pressure profiles behave realistically except a central singularity. It is shown that the energy conditions are satisfied in our study.
The equilibrium conditions through TOV equation  and stability criteria through Adiabatic index for the charged stellar structure study are investigated. Lastly, we have also provided a little review of the case with Reissner-Nordstrom spacetime as an exterior geometry for the matching condition. In both cases, the masses obey the Andreasson's limit $\sqrt{M} \leq \sqrt{R}/3+\sqrt{R/9+q^2/3R}$ requirement for a charged star. Conclusively, the results show that Bardeen model geometry provides more massive stellar objects as compared to usual Reissner-Nordstrom spacetime. In particular, the current study supports the existence of realistic massive structures like PSR J $1614-2230$.
\end{abstract}

\maketitle
%
%

\section{Introduction}

Compact star study has been an attractive topic of research in the recent past.
In fact, ``compact stars'' refer to a class of stellar objects like quark stars, white dwarfs, brown
dwarfs, neutron stars etc. Using some favorable equation of state (EoS) parameter can be helpful in getting realistic stellar structures.
MIT bag model EoS for the pressure-energy density relationship has been considered important to describe a fluid composed by quarks (up, down and strange) \cite{MIT1}. In particular, it has been used to investigate the stellar structure of compact stars in different situations \cite{MIT0}-\cite{3400}.
The stability of a stellar structure can be argued by investigating the Tolman-Oppenheimer-Volkoff (TOV) equations \cite{TOV1}-\cite{TOV3}.
An important feature of the TOV equation is the balancing act of some forces that actually describes that compact stars in a particular environment are physically viable. Moreover, the role of modified TOV equations to investigating compact stars structure in different alternative theories of gravity has been very important \cite{Ast2}-\cite{Ast3}.

Symmetries can play an important role to investigate the natural relation between geometry and
matter through the Einstein's equations. Among the well known symmetries, conformal
killing vectors (CKV's) can provide some better results as they provide a deeper insight into the spacetime geometry.
The CKV's are governed by the equation
\begin{eqnarray}\label{Killing}
\pounds_\xi g_{ij}=\phi~ g_{ij},
\end{eqnarray}
where for a four dimensional spacetime, $i,j=1,2,3,4$ and the quantity on the left hand side is known as the Lie
derivative of the metric tensor along the vector field $\xi$. In general, $\phi$ is an arbitrary function of radial coordinate $r$ and time $t$ (even in the static case) \cite{H14}.
It is worthwhile to mention here that Eq. (\ref{Killing}) generates homotheties when $\phi$ is constant and one obtain the Killing vectors when
$\phi=0$. Many interesting works from the literature suggest that compact stars can be modeled with the help of solutions admitting one parameter group of conformal motions. Herrera and his collaborators \cite{H10}-\cite{H13} were among the pioneers who gave the general treatment of the spheres admitting a one parameter group of conformal motions. Recent literature also provides some important results using CKV's \cite{Jamil}-\cite{Jamil1}. Mak and Harko \cite{MAK1} calculated an exact spherically symmetric solution describing the interior of a charged strange star assuming a one-parameter group of conformal motions. The problem of finding static spherically symmetric anisotropic compact star solutions have been addressed in general relativity using conformal motions \cite{512}. Esculpi and Aloma \cite{Esculpi} explored two new families of compact star solutions with charged anisotropic fluid admitting a one parameter group of conformal motions. Sharif and Waseem \cite{SW} investigated charged gravastars with conformal motion in $f(R,T)$ gravity. In another paper \cite{SW1}, Sharif and his co-author used the technique of conformal Killing motions to explore interior solutions for static spherically symmetric metric in the background of $f(G)$ gravity.

A black hole consists of a horizon having a singularity in it. But there may exist some black holes without any
singularity. Such black holes are termed as ``regular" black holes.
Bardeen was the first one who explored a black hole solution without singularity \cite{Bardeen}. Moreno
and Sarbach \cite{Moreno} found the stability properties
of the Bardeen black hole. The geodesic structures of test particles with Bardeen space-time have been analyzed by Zhou et al. \cite{Zhou}.
However, very few attempts have been made so far to study the charged compact structures using the interesting nature of Bardeen space-time under conformal motion. Therefore in this work, we are focussed to study the compact stars solutions using conformal motions and employing Bardeen geometry as an exterior spacetime. This will make the analysis interesting as Bardeen solution can be interpreted
as a gravitationally collapsed magnetic monopole arising from some specific case
of non-linear electrodynamics. A brief lay out of the paper is as follows: In section \textbf{II}, we provide some basic Einstein-Maxwell field equations formalism and include some discussion about CKV's and corresponding physical quantities. Section \textbf{III} is devoted to describe some important boundary and matching conditions using the Bardeen model as an exterior spacetime. Detailed discussions of some physical properties of relativistic spheres for the present study are reported in the fourth section. Last section provides a brief summary of the work and conclusive remarks.

\section{Field Equations: Conformal Motion Treatment}

We consider a spherically symmetric static distribution of a compact
star. In Schwarzschild coordinates, the space-time takes the following form
\begin{equation}\label{6}
ds^{2}=e^{\nu(r)}dt^{2}-e^{\lambda(r)}dr^2-r^{2}d\theta^{2}-r^2sin^{2}\theta d\phi^{2}.
\end{equation}
The source of energy-momentum tensor in the presence of charge is given as
\begin{equation}\label{4}
\mathcal{T}_{\chi\gamma}=(\rho+p)\upsilon_{\chi}\upsilon_{\gamma}-p\mathrm{g}_{\chi\gamma}+\frac{1}{4 \pi} (-\mathcal{F}^{\zeta \chi} \mathcal{F}_{\eta \chi} + \frac{1}{4} \delta^{\zeta}_{\eta} \mathcal{F}^{\chi \psi} \mathcal{F}_{\chi \psi} ),
\end{equation}
where $p$ is the pressure source, $\rho$ is an energy density source and $\mathcal{F}^{\zeta \chi}$ is the usual Maxwell's Stress Tensor.
The four velocity vector is denoted by $\upsilon_{\chi}$ and the radial four vector by $\xi_{\alpha}$, which satisfy the conditions
$\upsilon^{\alpha}=e^{\frac{-\nu}{2}}\delta^{\alpha}_{0}$, $\upsilon^{\alpha}\upsilon_{\alpha}=1.$
For the present analysis, we choose the MIT bag model EoS for the pressure-energy density relationship.
Such an EoS describes a fluid composed by quarks (up, down and strange) \cite{MIT1}. It has been used to investigate the
stellar structure of compact stars\cite{MIT2,MIT3}. MIT Eos model is given by
\begin{equation}\label{MIT}
p = a(\rho-4B).
\end{equation}
The parameter $B$ is known as the bag constant while the constant $a$ may be assumed to be equal to $1/3$  \cite{MIT4}.
Assuming the gravitational units $c=1=G$, Einstein-Maxwell's Field equations are given by
\begin{eqnarray}\nonumber
R_{\mu\nu}-\frac{1}{2} Rg_{\mu\nu} = 8\pi T_{\mu\nu},\\\label{600}
\mathcal{F}^{\eta\zeta}_{;\zeta}=-4 \pi j^\eta,~~~~~~~~~~~~~~~~
\mathcal{F}_{[\beta\zeta;\eta]}&=&0,
\end{eqnarray}
where $j^\eta$ is the electromagnetic four current vector. Maxwell's stress tensor and electromagnetic four current vector are defined as
\begin{eqnarray}\label{12}
\mathcal{F}^{\eta \zeta}=A_{\zeta,\eta}-A_{\eta, \zeta},~~~~~~~
j^\eta=\sigma \nu^{\eta},
\end{eqnarray}
respectively. Here $A$ being the magnetic four potential and $\sigma$ symbolize as charge density.
The only non vanishing component in the static spherically symmetric system is $J^{0}$. The Einstein-Maxwell tensor only contains the non vanishing component $\mathcal{F}^{01}$ and is defined as
\begin{eqnarray}\label{13}
\mathcal{F}^{01}=-\mathcal{F}^{10}=\frac{q}{r^2} e^{-(\frac{\nu+\lambda}{2})}.
\end{eqnarray}
The term electric field intensity $E$ can be expressed as
\begin{eqnarray}
~~~~~~~~~~E^2=-\mathcal{F}^{01}\mathcal{F}_{10}=\frac{q^2}{ r^4},~~~~~~~~~~\quad\quad~~~~~~\label{19}
\end{eqnarray}
where the term $q$ represent charge inside the spherical stellar system of radial function $r$ is given by
\begin{eqnarray}\label{14}
q=4\pi\int_{0}^{r} 4 \sigma \rho^2 e^{(\frac{\lambda}{2})} d\rho.
\end{eqnarray}
Now for the spacetime (\ref{6}) along with the matter distribution (\ref{4}), Einstein-Maxwell's Field equations (\ref{600}) turn out to be
\begin{eqnarray}\label{F1}
8\pi\rho+E^2&=&\frac{1}{ e^{\lambda}r^2}\big(\lambda'r+e^{\lambda}-1\big),
\\\label{F2}
8\pi p-E^2&=&\frac{1}{ e^{\lambda}r^2}\big(\nu'r-e^{\lambda}+1\big),~\\\label{F3}
8\pi p+E^2&=&\frac{1}{ e^{\lambda}}\big(\frac{{\nu'}^{2}}{4}+\frac{\nu''}{2}-\frac{\nu'\lambda'}{4}+\frac{\nu'}{2r}-\frac{\lambda'}{2r}\big),\\\label{F4}
 \sigma&=& \frac{e^{-\lambda/2}}{4\pi r^2}(r^2E)'.
\end{eqnarray}
Now applying equation (\ref{Killing}) on the spacetime (\ref{6}), we get the following conformal Killing equations \cite{H13}
\begin{eqnarray}\label{Killing1}
\xi^1 \nu'=\phi,~~~~~ \xi^4=L,~~~~~\xi^1=\frac{\phi~r}{2},~~~~~\xi^1 \lambda'+2{\xi^1}_1=\phi,
\end{eqnarray}
which further yield a simultaneous solution as follows
\begin{eqnarray}\label{Killing2}
e^\nu=A^2r^2,~~~~~ e^\lambda=\big(\frac{S}{\phi}\big)^2,~~~~~\xi^i=L~{\delta^i_4} +\big(\frac{r\phi}{2}\big){\delta^i_1},
\end{eqnarray}
where $A$, $L$ and $S$ are arbitrary constants. For the sake of simplicity, we introduce a new variable $N(r)=\big(\frac{\phi}{S}\big)^2$ by assuming
$\phi$ as a function of radial coordinate $r$ only. Manipulating field equations (\ref{F1}-\ref{F4}), MIT model (\ref{MIT}) and the solution obtained through conformal motion (\ref{Killing2}), we get
\begin{eqnarray}\label{DE1}
\frac{dN}{dr}=-\frac{2N}{r}+\frac{2}{3r}-\frac{4Br}{3},
\end{eqnarray}
whose straightforward solution is obtained as
\begin{eqnarray}\label{DE2}
N=\frac{1}{3}(1-Br^2)+\frac{C}{r^2},
\end{eqnarray}
where $C$ is an integration constant. It is worthwhile to mention here that exactly the same solution has been reported by Mak and Harko \cite{MaK}. However, they assumed the integration constant $C=0$ to analyze compact stars with usual Reissner-Nordstrom geometry for matching conditions using conformal symmetries. In this work, we consider $C\neq0$ and use Bardeen geometry as an exterior spacetime to study compact stars. Thus, we obtain the following explicit exact solution describing the interior of a charged strange star as
\begin{eqnarray}\label{Killing5}
e^\nu=A^2r^2,~~~~~ e^\lambda=\frac{3r^2}{3C+r^2(1-Br^2)},~~~~~\rho=\frac{2 B r^4+6 C+r^2}{16 \pi  r^4},~~~~P=\frac{r^2-6 B r^4+6 C}{48 \pi  r^4}.
\end{eqnarray}
Moreover, the electric field intensity $E$ turn out to be
\begin{eqnarray}
E^2=\frac{r^2-12 C}{6 r^4}.
\end{eqnarray}
It is interesting to notice that these explicit solutions involve three arbitrary parameters including the Bag constant. However, the parameter $A$ will have no role in further analysis but $B$ and $C$ are very important for constructing compact star solutions. Moreover, the physical parameters do have a singularity at center which is due to the use of conformal symmetries. In fact, this is the only drawback that the central singularity in the physical parameters can not be avoided in this formalism \cite{Jamil,MAK1,Rahman123,AbbasShahzad}. However, the solutions can be studied to describe the envelope region of a star
in a core-envelope type model.
Now we discuss some suitable boundary conditions to match the solutions of interior spacetime.

\section{Boundary and Matching Conditions}

For a well behaved viable solution, the following conditions must be hold.
\begin{itemize}
\item There must be a region $r=R_b$, i.e. the surface of the star, where the pressure is zero $P(R_b)=0$. Moreover, the metric functions should be finite and bounded, must not have singularities throughout the star and also at boundary, i.e for $0\le r \leq R_b$.
\item The pressure and density functions must be positive and monotonically decreasing and at the center their value must be maximum i.e.,
\begin{equation*}
P(0) > 0,~~~\frac{dP}{dr}{\mid}_{r=0} = 0,~~~\frac{d^{2}P}{dr^{2}}{\mid}_{r=0} < 0,~~~
\rho(0) > 0, ~~~~\frac{d\rho}{dr}{\mid}_{r=0} = 0,~~~~\frac{d^{2}\rho}{dr^{2}}{\mid}_{r=0} < 0,
\end{equation*}
and for $r \neq 0$, ${\rho}' < 0$ and $P' < 0$. Moreover, the energy conditions must be satisfied.
\item The causality condition must not be violated, i.e. the magnitude of speed of sound must be less than the speed of light, i.e.
$0\leq v^2 = \frac{\partial P(\rho)}{\partial\rho}\leq 1$.
\item   The solution must be stable, for stability adiabatic index is required to be satisfied,
$\gamma= \frac{\rho + P}{P}v^{2}>\frac{4}{3}$.
\item The mass function must obey the Andreasson's limit $\sqrt{M} \leq \frac{\sqrt{R_b}}{3}+\sqrt{\frac{R_b}{9}+\frac{q^2}{3R_b}}$ requirement for a charged star \cite{Anderson}.
\end{itemize}
For further analysis, we propose Bardeen model to describe as an exterior
spacetime given by \cite{Bardeen}
\begin{equation}\label{bardeen}
ds^{2}=h(r)dt^{2}-{h(r)}^{-1}dr^2-r^{2}d\theta^{2}-r^2sin^{2}\theta d\phi^{2},
\end{equation}
where $h(r)=1 -\frac{2Mr^2}{({q^2}+{r^2})^{\frac{3}{2}}}$. The interesting feature of this model is that it has been shown that the Bardeen black hole can be interpreted as a gravitationally collapsed magnetic monopole arising from some specific case
of non-linear electrodynamics \cite{Garcia}. In particular, Bardeen black hole has provided interesting results on many occasions \cite{Fernando1}-\cite{Ulhoa}. It is worth mentioning that the spacetime asymptotically behaves as
\begin{equation}\label{29}
h(r) = 1-\frac{2M}{r}+\frac{3Mq^2}{r^3} + O(\frac{1}{r^5}).
\end{equation}
It is interesting to see that the term $1/r$ in Eq. (\ref{29}) shows that the parameter $M$ associated with the usual mass of the stellar configuration. However, the next term involving $1/r^3$ makes the case more interesting and different from the usual Reissner–
Nordstrom solution \cite{Nordstrom}. Thus for the present analysis, we assume $h(r)\approx 1-\frac{2M}{r}+\frac{3Mq^2}{r^3}$.
Imposing the continuity condition for the metric potentials on boundary, matching equations are as follows:
\begin{eqnarray}\label{M1}
1-\frac{2M}{{R_b}}+\frac{3Mq^2}{{R_b}^3}=A^2 r^2,~~~~
(1-\frac{2M}{{R_b}}+\frac{3Mq^2}{{R_b}^3})^{-1}=\frac{3 r^2}{r^2 \left(1-B r^2\right)+3 C},~~~~~
p_r(r = R_b) = 0.
\end{eqnarray}
These matching conditions (\ref{M1}) can be solved simultaneously to obtain explicit expressions for the parameters $A$, $B$ and $C$ as
\begin{eqnarray}\label{M11}
A=\pm\sqrt{\frac{M+r}{r^2 (9 M+r)}},~~~~
B=\frac{5 r-3 M}{4 r^2 (9 M+r)},~~~~
C=\frac{r^2 (13 r-27 M)}{12 (9 M+r)}.
\end{eqnarray}
Now we investigate the suitable solution in the light of above mentioned conditions.
In the next section, we provide the important physical analysis using the free parameters $R_b,~ B$ and $C$.
\begin{center}
\begin{table}[h]
\caption{\label{tab1}{Approximated values of $R_b,\; M,$ and $C$ with $B=0.002$.}}
\vspace{0.3cm}\begin{tabular}{|c|c|c|c|c|c|c|}
\hline
$R_b$                \;\;\;\;\;\;\;\;\;\;\;\;& $M(M_{\odot})$ \;\;\;\;& $C$\;\;\;\;& $\sqrt{M}$ \;\;\;\;& $\sqrt{M}/3+\sqrt{R_b/9+q^2/3R_b}$   \\

 \hline
9.12904               \;\;\;\;\;\;\;\;\;\;\;\;&  2.97979661        \;\;\;\;\;\;\;\;\;\;\;\;&0.001  \;\;\;\;\;\;\;&1.72621 \;\;\;\;\;\;\;\;&2.24061  \\
 \hline
9.14507               \;\;\;\;\;\;\;\;\;\;\;\;&  2.97644068        \;\;\;\;\;\;\;\;\;\;\;\;&0.050  \;\;\;\;\;\;\;&1.72524 \;\;\;\;\;\;\;&2.24113 \\
  \hline
9.20909               \;\;\;\;\;\;\;\;\;\;\;\;&  2.96298998      \;\;\;\;\;\;\;\;\;\;\;\;&0.250    \;\;\;\;\;\;\;\;&1.72133 \;\;\;\;\;\;\;\;&2.24311\\
 \hline
9.30119               \;\;\;\;\;\;\;\;\;\;\;\;&  2.94349831        \;\;\;\;\;\;\;\;\;\;\;\;&0.550    \;\;\;\;\;\;\;&1.71566 \;\;\;\;\;\;\;&2.24573\\
\hline
9.36021               \;\;\;\;\;\;\;\;\;\;\;\;&  2.93093559       \;\;\;\;\;\;\;\;\;\;\;\;&0.750    \;\;\;\;\;\;\;&1.71200 \;\;\;\;\;\;\;&2.24726 \\
\hline
\end{tabular}
\end{table}
\end{center}

\section{Qualitative Analysis}

It is evident from Eq. (\ref{Killing5}) that the obtained solutions through conformal symmetries for the metric potentials are
finite, bounded and do not possess any singularity throughout the star and also at boundary, i.e for $0\le r \leq R_b$.
It is worthwhile to mention here that for all the graphical analysis, the value of Bag constant has been fixed to be $B=0.002$, which less than the predicted value of Bag constant $0.98<B<1.52$ in the units of $B_0=60 MeV/fm^3$ \cite{MIT4}, but otherwise we might not get some realistic results for analyzing compact stars in this study. Perhaps, this is due to the use of solutions through conformal symmetries. Moreover, the constant $C$, which was considered to be zero in some of previous work \cite{MaK}, plays a very important role in analyzing the different physical quantities for quark star discussions in present study.
The parameter $A$ has no contribution in the current analysis.

The evolution of energy density and pressure functions is exhibited in Fig. \textbf{1}. Both energy density and pressure profiles behave realistically, except an unavoidable central singularity.
The $\rho$ is seen positive, decreasing to a minimum value at $r=R_{b}$ while pressure approaches to zero near the boundary of the star. The concave up growth of the curve for energy density and pressure is due to conformal symmetry motions and the presence of electric charge. The analysis is done with $R_{b}=9.36021,~9.30119,~9.20909,~9.14507,~9.12904$. Magnified views are also inserted in Fig. \textbf{1}, which show that energy density and pressure profiles  are in order of $10^{15} g/cm^3$ and $10^{35} dyne/cm ^2$ which are in realistic range for a neutron or quark star.
Further details are shown in table \textbf{I}.
The derivatives of energy density and pressure functions with respect to radial coordinate, i.e., $d\rho/dr$ and $dp/dr$ are observed negative for the current study (Fig. \textbf{2}). The negativity in gradients shows that our obtained solutions are physically acceptable.
\begin{figure}
\centering \epsfig{file=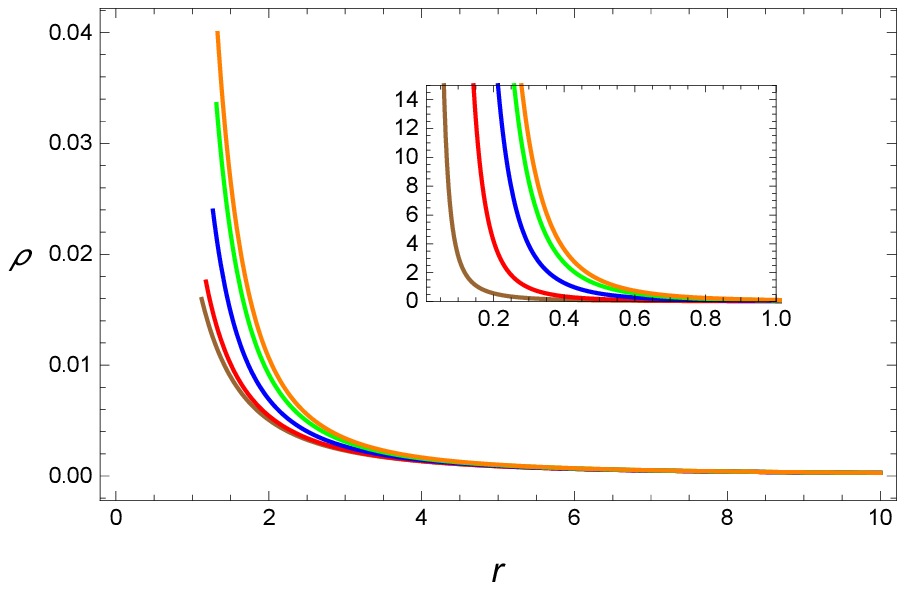, width=.48\linewidth,
height=2.2in}\epsfig{file=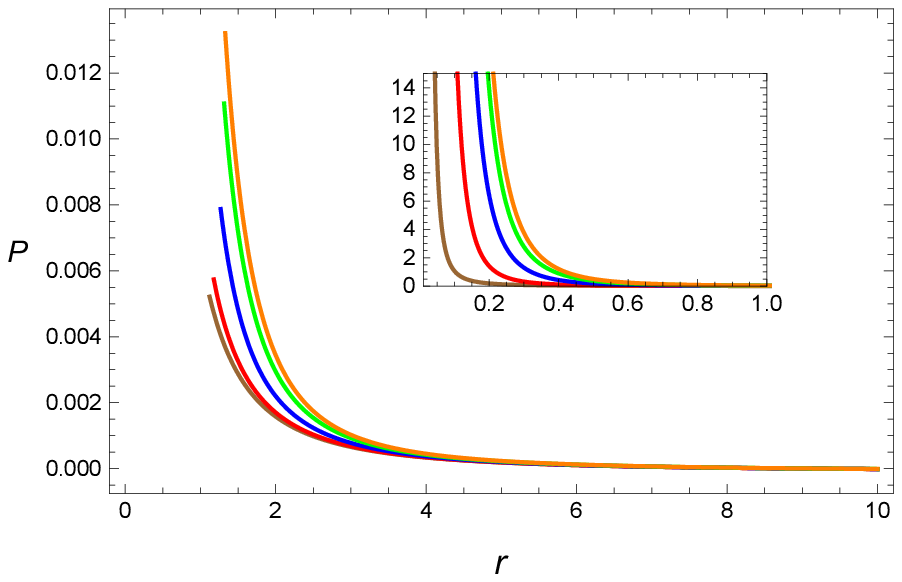, width=.48\linewidth,
height=2.2in}\caption{\label{Fig.1} Behavior of energy density and pressure $[MeV/fm^3]$ vs radial coordinate $(Km)$, with $R_{b}=9.36021(\textcolor{orange}{\bigstar})$, $R_{b}=9.30119(\textcolor{green}{\bigstar})$, $R_{b}=9.20909(\textcolor{blue}{\bigstar})$, $R_{b}=9.14507(\textcolor{red}{\bigstar})$, $R_{b}=9.12904(\textcolor{brown}{\bigstar})$.}
\end{figure}
\begin{figure}
\centering \epsfig{file=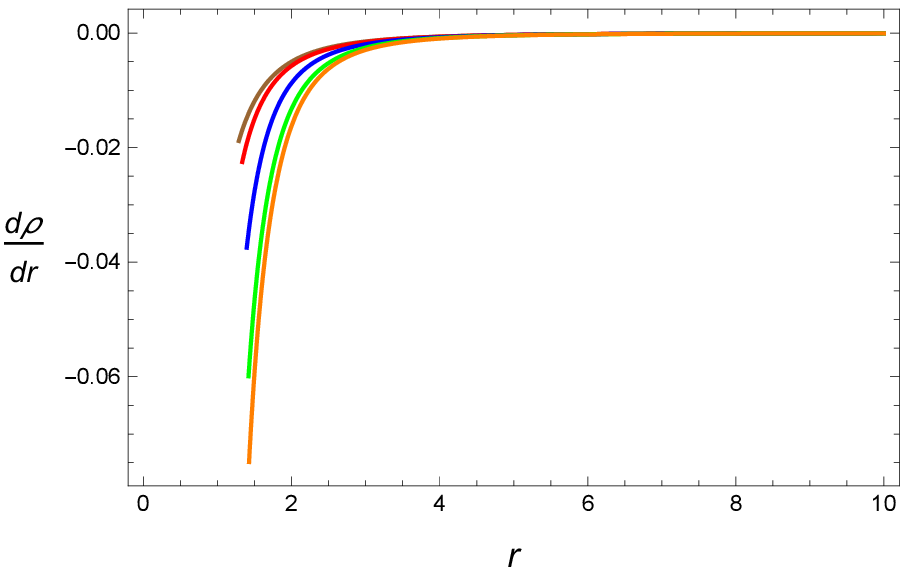, width=.48\linewidth,
height=2.2in}\epsfig{file=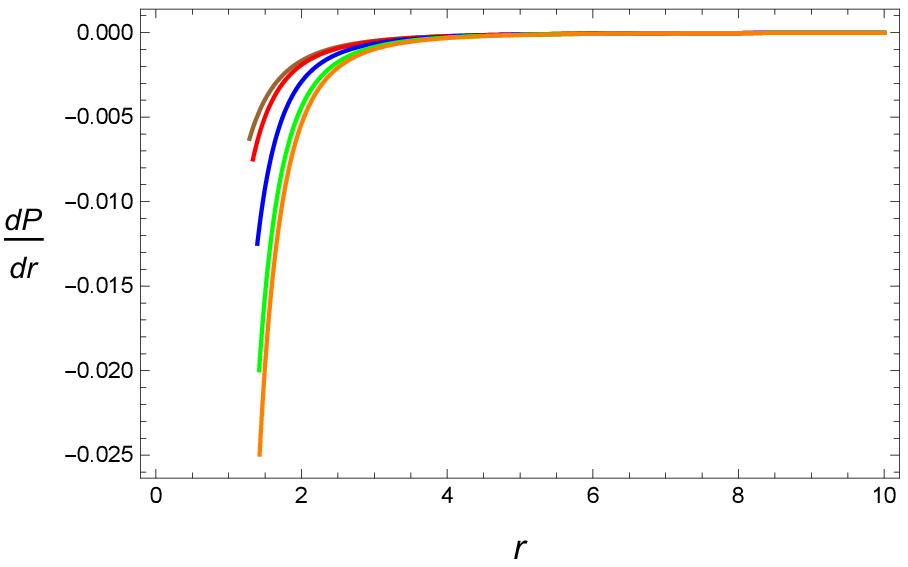, width=.48\linewidth,
height=2.2in}\caption{\label{Fig.2} Behavior of derivatives of energy density and pressure functions, with $R_{b}=9.36021(\textcolor{orange}{\bigstar})$, $R_{b}=9.30119(\textcolor{green}{\bigstar})$, $R_{b}=9.20909(\textcolor{blue}{\bigstar})$, $R_{b}=9.14507(\textcolor{red}{\bigstar})$, $R_{b}=9.12904(\textcolor{brown}{\bigstar})$.}
\end{figure}
\begin{figure}
\centering \epsfig{file=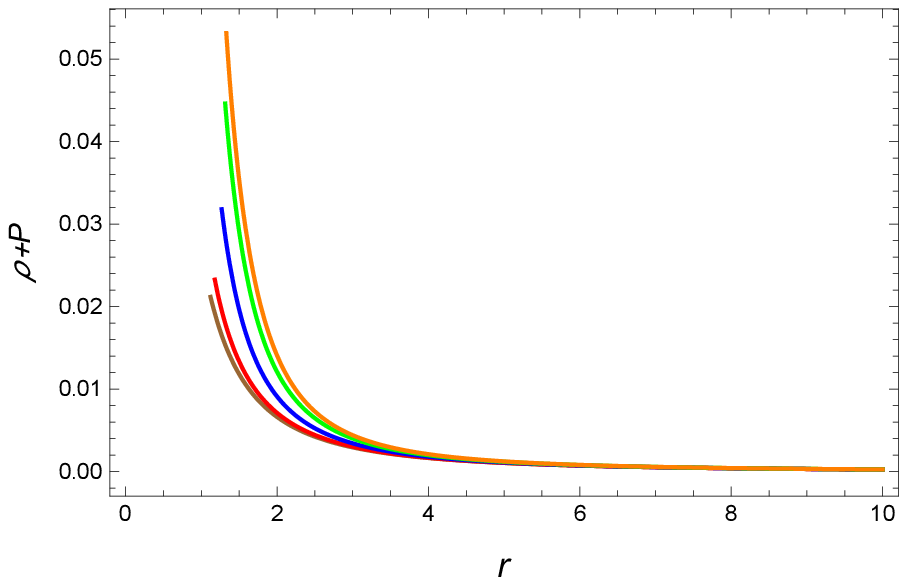, width=.48\linewidth,
height=2.2in}\epsfig{file=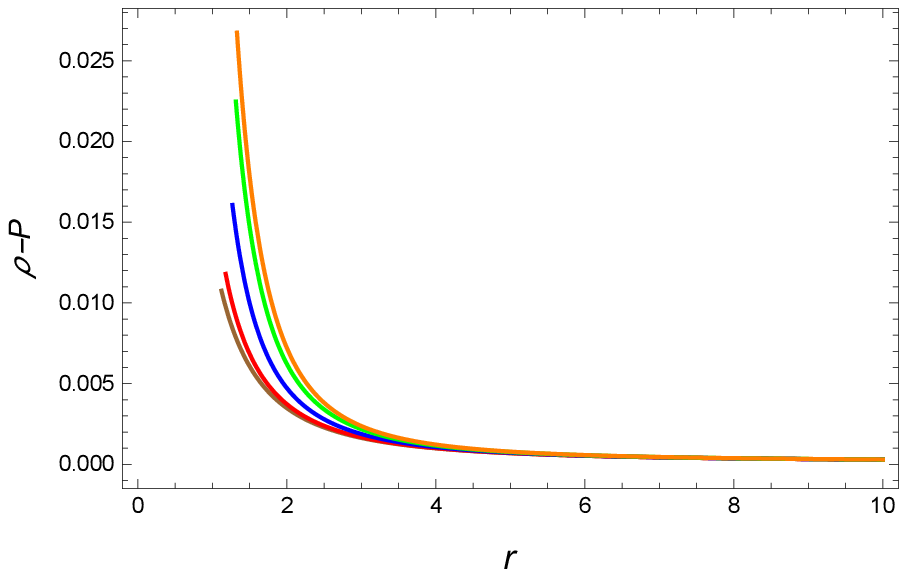, width=.48\linewidth,
height=2.2in}\caption{\label{Fig.3} Behavior of energy conditions ($WEC$ and $DEC$), with $R_{b}=9.36021(\textcolor{orange}{\bigstar})$, $R_{b}=9.30119(\textcolor{green}{\bigstar})$, $R_{b}=9.20909(\textcolor{blue}{\bigstar})$, $R_{b}=9.14507(\textcolor{red}{\bigstar})$, $R_{b}=9.12904(\textcolor{brown}{\bigstar})$.}
\end{figure}

\subsection{Energy Conditions}

The study of energy conditions has many significant applications in general relativistic and cosmological contexts. It is interesting that the Hawking-Penrose singularity theorems and the validity of the second law of black hole thermodynamics can be studied using the energy conditions \cite{EC2}. In relativistic cosmology, many interesting results are discussed with the help of energy conditions \cite{EC3}-\cite{EC11}. Five kinds of energy bounds are found in literature as described below:
\begin{itemize}
  \item Trace energy condition $(TEC)$, now abandoned,
  \item Null energy condition $(NEC)$,
  \item Weak energy condition $(WEC)$,
  \item Strong energy condition $(SEC)$,
  \item Dominant energy condition $(DEC)$.
\end{itemize}
The trace energy condition suggests that the
trace of the energy-momentum tensor should always be negative (or positive depending on metric conventions). This condition was popular among the researchers during the decade of $1960$. However, once it was shown that stiff EoS, such as those which are appropriate for neutron stars, violate the TEC \cite{EC,EC1}. Thus the study of this energy condition was not further encouraged and it is now completely abandoned, in fact no longer cited in the literature.
However, the remaining four energy bounds are assumed as necessary features for compact stars discussion in the background of general relativity and also other modified theories of gravity. For an acceptable stellar structure model these energy conditions should be satisfied, i.e. $NEC$ $\rho + P\geq 0$, $WEC$  $\rho\geq 0$ and $\rho + P\geq 0$, $SEC$ $\rho + 3P\geq 0$ and $\rho + P\geq 0$,
and $DEC$ $\rho\geq 0$ and $\rho \pm P\geq 0$. It is evident from Figs. \textbf{1, 3, 4} that all these energy conditions are satisfied in our case. Moreover for our curiosity we have also investigated $TEC$ which turns out to be constant, i.e. $\frac{B}{2 \pi }>0$, which should turn out to be negative but not the case with neutron stars as already discussed.
\begin{figure}
\centering \epsfig{file=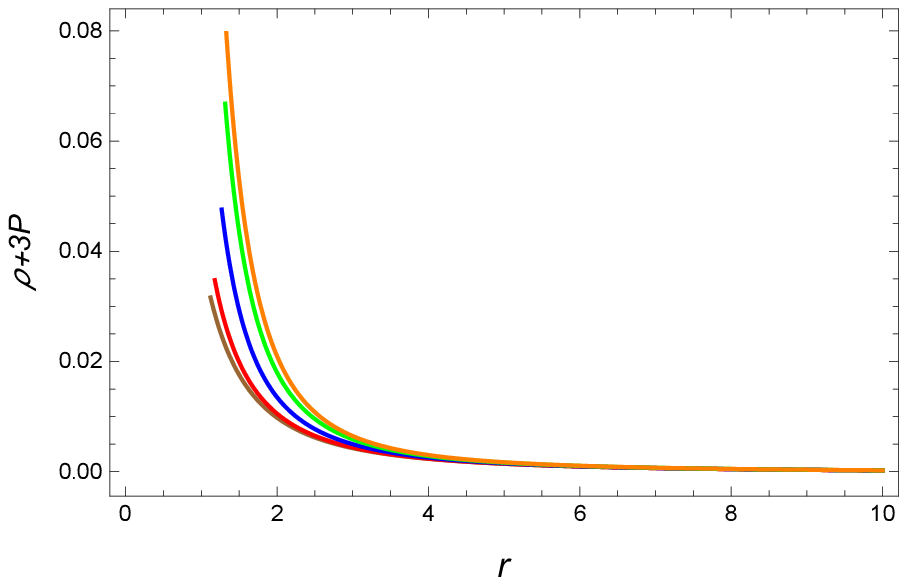, width=.48\linewidth,
height=2.2in}\epsfig{file=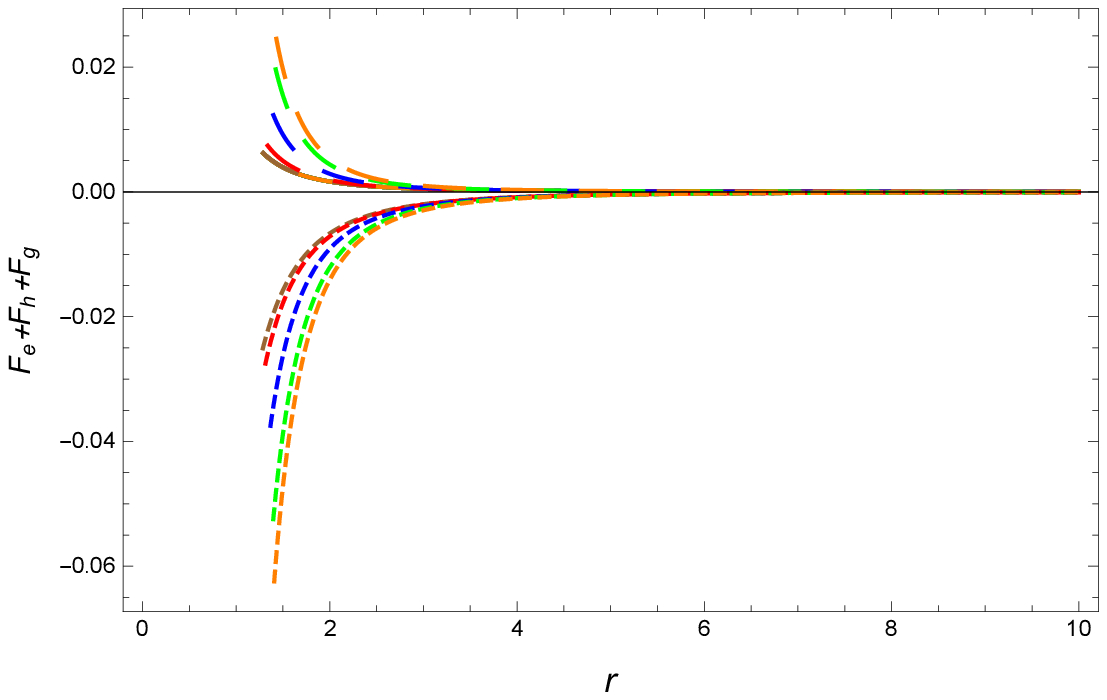, width=.48\linewidth,
height=2.2in}\caption{\label{Fig.3} Behavior of energy condition ($SEC$) and TOV equation, with $R_{b}=9.36021(\textcolor{orange}{\bigstar})$, $R_{b}=9.30119(\textcolor{green}{\bigstar})$, $R_{b}=9.20909(\textcolor{blue}{\bigstar})$, $R_{b}=9.14507(\textcolor{red}{\bigstar})$, $R_{b}=9.12904(\textcolor{brown}{\bigstar})$. In right plot, Non-dashed curves to represent hydrostatic force ($F_h$), dashed curves to represent electric force ($F_e$) and small dashed curves to represent the gravitational force ($F_g$).}
\end{figure}
\subsection{Equilibrium and Stability Analysis}

In this sub-section, we discuss the equilibrium conditions for the current charged stellar structure study in the background of Bardeen geometry admitting conformal motions. It is worthwhile to mention here that the TOV equation is very important to analyze the equilibrium conditions for a stellar structure. The TOV equation mainly constrains the structure of a static spherically symmetric object which is in static gravitational equilibrium. In particular, it becomes interesting to see how gravitational and other fluid forces behave with the increasing electrostatic repulsion towards the
boundary of star. Thus, a generalized charged TOV equation can be described as \cite{TOV1,TOV2}
\begin{equation}\label{36}
-\frac{d p }{dr}-\frac{M_G(r)}{r^2}(\rho+p)e^{\frac{\lambda (r)-\nu(r)}{2}}+ \sigma(r)E(r)e^{\frac{\lambda (r)}{2}}=0,
\end{equation}
where $M_G(r)$ is knows as the effective gravitational mass within a charged quark. The expression for the effective gravitational mass takes the form \cite{TOV3}
\begin{equation}
M_G=\frac{1}{2}r^2 \nu'e^{\frac{\nu-\lambda}{2}},
\end{equation}
and consequently Eq. (\ref{36}) becomes
\begin{equation}
-\frac{d p }{dr}-\frac{\nu'}{2}(\rho+p)+ \sigma E e^{\frac{\lambda }{2}}=0.
\end{equation}
This equation can be viewed as
\begin{equation}\label{37}
{F}_{\mathrm{h}}+{F}_{\mathrm{g}}+{F}_{\mathrm{e}}=0,
\end{equation}
where
\begin{itemize}
   \item ${F}_{\mathrm{h}}=-\frac{d p}{dr}$ defines the hydrostatic force,
  \item ${F}_{\mathrm{g}}=-\frac{\nu'}{2}(\rho+p)$ reveals the gravitational force,
  \item ${F}_{\mathrm{e}}=\sigma E e^{\frac{\lambda }{2}}$ mentions the electric force.
\end{itemize}
The interesting behavior of the hydrostatic, gravitational and electric forces can be seen balanced, which is shown in the right plot of Fig. \textbf{4}.
The non-dashed and the dashed curves above the axis represent hydrostatic and electric forces while the small dashed curves below the axis are to represent the gravitational force.
The balancing trend of all the above forces, i.e., ${F}_{\mathrm{h}}$, ${F}_{\mathrm{g}}$ and ${F}_{\mathrm{e}}$ indicates that our obtained electrically charged and conformal motion solutions for the compact stars are stable and physically acceptable.\\\\
The stability analysis of a compact star study is also an important feature. Hillebrandt and Steinmetz \cite{Hil}, discussed an important parameter known as Adiabatic index, given by
\begin{equation}\label{39}
\gamma= \frac{\rho + P}{P}v^{2},
\end{equation}
where $v^{2}$ is to denote magnitude of speed of sound. In fact, another interesting constraint is that the causality condition must not be violated, i.e. the magnitude of speed of sound must be less than the speed of light, i.e.
$0\leq v^2 = \frac{\partial P}{\partial\rho}\leq 1$. In this study, we obtain $v^2=1/3$ which is in a good agreement with the requirement to justify the causality condition. Moreover, the behavior of Adiabatic index $\gamma_{r}$ is shown in figure \textbf{5}. It is clearly shown that all the curves involved remain in a zone $>4/3$. Thus the results with $\gamma_{r}>4/3$ indicate that solutions with conformal motion with Bardeen model as an exterior geometry of star support stability criteria through Adiabatic index.

\subsection{Comparison with Reissner-Nordstrom Case}

Now we provide a little review of the case with Reissner-Nordstrom spacetime as an exterior geometry for the matching condition. This will also enable us to provide some comparison with our study. The Reissner-Nordstrom spacetime is given by
\begin{equation}\label{RN}
ds^{2}=(1-\frac{2M}{r}+\frac{q^2}{r^2})dt^{2}-(1-\frac{2M}{r}+\frac{q^2}{r^2})^{-1}dr^2-r^{2}d\theta^{2}-r^2sin^{2}\theta d\phi^{2}.
\end{equation}
Similarly as in the previous case, we impose the continuity condition for the metric potentials on boundary to obtain the following matching equations:
\begin{eqnarray}\label{M1}
1-\frac{2M}{{R_b}}+\frac{q^2}{{R_b}^2}=A^2 r^2,~~~~
(1-\frac{2M}{{R_b}}+\frac{q^2}{{R_b}^2})^{-1}=\frac{3 r^2}{r^2 \left(1-B r^2\right)+3 C},~~~~~
p_r(r = R_b) = 0.
\end{eqnarray}
These matching conditions (\ref{M1}) can be solved simultaneously to obtain explicit expressions for the parameters $A$, $B$ and $C$ as
\begin{eqnarray}\label{M11}
A=\pm\sqrt{\frac{{r-M}}{{2} r^{3}}},~~~~
B=\frac{2 r-3 M}{4 r^3},~~~~
C=\frac{1}{12} r (4 r-9 M).
\end{eqnarray}
\begin{figure}
\centering \epsfig{file=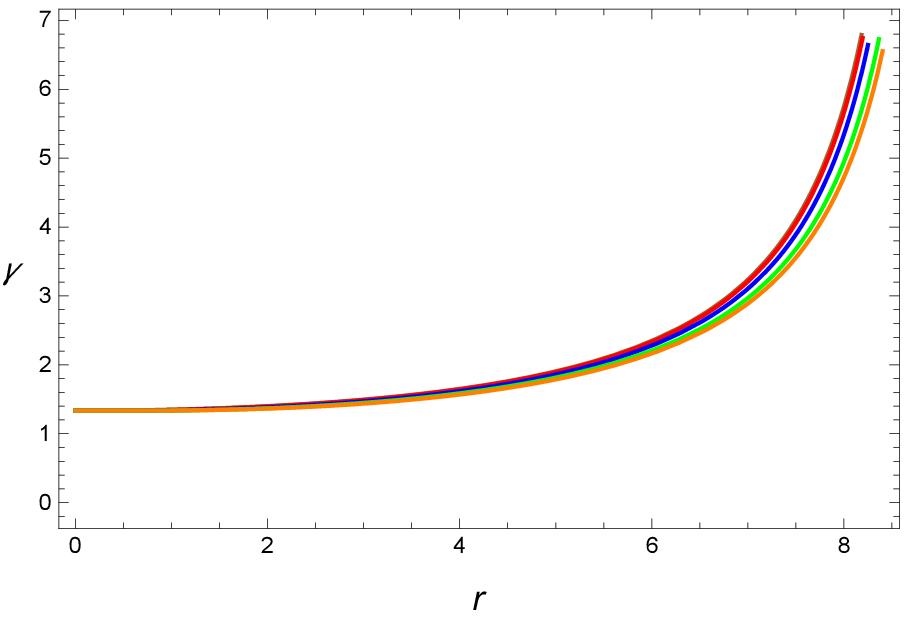, width=.48\linewidth,
height=2.2in},\caption{\label{Fig.3} Behavior of Adiabatic Index, with $R_{b}=9.36021(\textcolor{orange}{\bigstar})$, $R_{b}=9.30119(\textcolor{green}{\bigstar})$, $R_{b}=9.20909(\textcolor{blue}{\bigstar})$, $R_{b}=9.14507(\textcolor{red}{\bigstar})$, $R_{b}=9.12904(\textcolor{brown}{\bigstar})$.}
\end{figure}
\begin{center}
\begin{table}[h]
\caption{\label{tab1}{Approximated values of $R_b,\; M,$ and $C$ with $B=0.002$ for RN Case.}}
\vspace{0.3cm}\begin{tabular}{|c|c|c|c|c|c|c|}
\hline
$R_b$                \;\;\;\;\;\;\;\;\;\;\;\;& $M(M_{\odot})$ \;\;\;\;& $C$ \;\;\;\;& $\sqrt{M}$ \;\;\;\;& $\sqrt{M}/3+\sqrt{R_b/9+q^2/3R_b}$   \\

 \hline
9.12904               \;\;\;\;\;\;\;\;\;\;\;\;&  2.75064407        \;\;\;\;\;\;\;\;\;\;\;\;&0.001  \;\;\;\;\;\;\;&1.65850658 \;\;\;\;\;\;\;\;&2.24061  \\
 \hline
9.14507               \;\;\;\;\;\;\;\;\;\;\;\;&  2.75063051        \;\;\;\;\;\;\;\;\;\;\;\;&0.050   \;\;\;\;\;\;\;&1.65850249 \;\;\;\;\;\;\;&2.24113\\
  \hline
9.20909               \;\;\;\;\;\;\;\;\;\;\;\;&  2.75032542      \;\;\;\;\;\;\;\;\;\;\;\;&0.250    \;\;\;\;\;\;\;\;&1.65841051 \;\;\;\;\;\;\;\;&2.24311\\
 \hline
9.30119               \;\;\;\;\;\;\;\;\;\;\;\;&  2.74916610        \;\;\;\;\;\;\;\;\;\;\;\;&0.550    \;\;\;\;\;\;\;&1.65806095 \;\;\;\;\;\;\;&2.24573\\
\hline
9.36021               \;\;\;\;\;\;\;\;\;\;\;\;&  2.74797288      \;\;\;\;\;\;\;\;\;\;\;\;&0.750    \;\;\;\;\;\;\;&1.65770708 \;\;\;\;\;\;\;&2.24726\\
\hline
\end{tabular}
\end{table}
\end{center}
Now for a comparative analysis, we investigate the solutions in the light of above mentioned conditions.
As a first step we keep the values of parameters $R_b,~ B$ and $C$ same as in the Bardeen model case, so that we may get an idea what happens to the fourth parameter $M$. The results are tabulated in Table \textbf{II}. It is interesting to notice that the values of radii remain same as in the Bardeen case, however, decrease in the values of mass is evident in Table \textbf{II} as compared to those in the case of Bardeen model (see Table \textbf{I}). Moreover, the variation corresponding to given radii is so small that we have to increase the number of decimal place accuracy to see a difference. As a second step, when we try to recover the same mass in the case of Bardeen model, we do not get realistic results. For example, for $(M,R_b)=(2.75064407,9.12904)$ we obtain both $B$ and $C$ negative $(B=-3.0429,~C=-253.594)$, providing a negative profile for energy density. Therefore, the results show that Bardeen model case provides more massive stellar objects as compared to usual RN case. In case of both Bardeen and RN models, the masses obey the Andreasson's limit $\sqrt{M} \leq \frac{\sqrt{R_b}}{3}+\sqrt{\frac{R_b}{9}+\frac{q^2}{3R_b}}$ requirement for a charged star \cite{Anderson}, (kindly see Tables \textbf{I} and \textbf{II}).

\subsection{Comparison with Some Realistic Stellar Structure}

This subsection is devoted for some comparison with realistic structure. For this purpose, we choose the well-known model PSR J $1614-2230$ \cite{a1}. At the time of its discovery, the mass was observed to be $1.97 \pm 0.04 M_{\odot}$ and the radius as $13\pm 2~ km$. This observed mass made PSR J $1614–2230$ the most massive known neutron star at the time of its discovery, and ruled out many neutron star equations of state that included exotic matter such as hyperons and kaon condensates. The M-R diagram in the light of present study for PSR J $1614-2230$ is shown in Fig. \textbf{6}. It is important to notice that realistic range of mass of PSR J $1614-2230$ depends upon the value of constant $C$. The higher values do not favor the mass of PSR J $1614-2230$ as in observed range. However, the smaller values (kindly see brown and red colored curves in Fig. \textbf{6}) exactly correspond to the masses in favorable domain. In fact, it is interesting to notice that the effects of electric charge with Bardeen geometry actually may improve the net mass of the structure. Thus more massive stellar structures can be theorized in the context of this study with conformal motion as compared to previous work \cite{a2}. Moreover, the Andreasson's limit requirements are also met in this case (though exact values are not presented here in any tabular form).

\section{Outlook}

This work is devoted to discuss the solutions of Einstein-Maxwell's field equations for compact stars study in the background of conformal motions. For this purpose, we have considered perfect fluid source of matter with electric charge. For the present analysis, we choose MIT bag model EoS for the pressure-energy density relationship. Further, CKV's are used to investigate the appropriate forms for gravitational metric coefficients. In fact, MIT bag model EoS plays an important role in developing a differential equation using field equations and solutions through conformal motion. This provides us with a fundamental mathematical background for the further analysis. It would be worthwhile to mention here that we do not assume the integration constant equal to zero which is obtained while solving the differential equation. In fact, this integration constant plays a key role in our present compact star study.
Further, we impose the boundary conditions, by choosing the Bardeen model to describe as an exterior spacetime \cite{Bardeen}. The Bardeen model provides the analysis in an interesting and unique way with good accuracy. Moreover, Bardeen solution can be interpreted as a gravitationally collapsed magnetic monopole arising from some specific case of non-linear electrodynamics. A brief qualitative analysis of the current work is itemized below:\\
\begin{figure}
\centering \epsfig{file=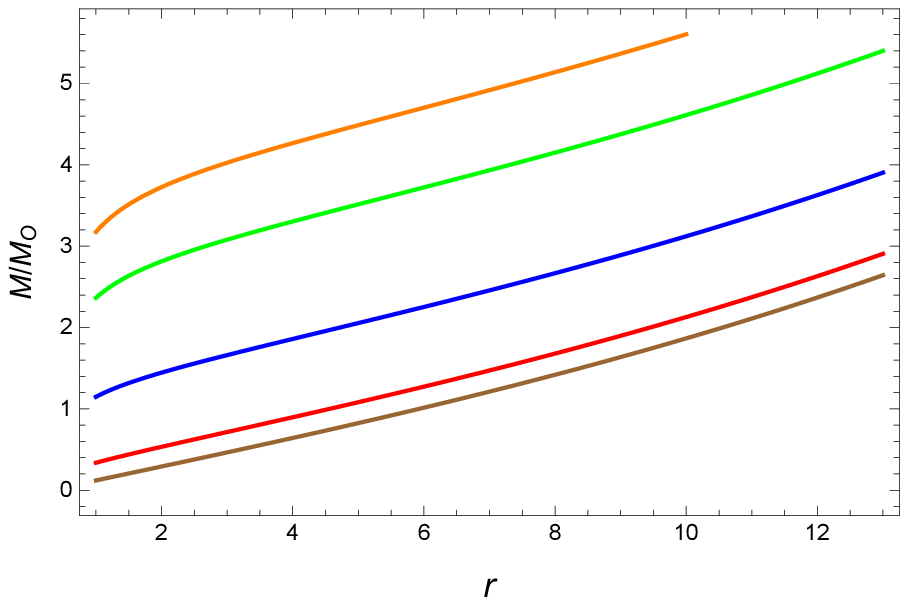, width=.48\linewidth,
height=2.2in},\caption{\label{Fig.3} Mass-Radius Diagram for PSR J $1614-2230$, with $B=0.002$ and $C=0.001(\textcolor{brown}{\bigstar})$, $C=0.05(\textcolor{red}{\bigstar})$, $C=0.25(\textcolor{blue}{\bigstar})$, $C=0.55(\textcolor{green}{\bigstar})$, $C=0.75(\textcolor{orange}{\bigstar})$.}
\end{figure}
\begin{itemize}
  \item The obtained solutions through conformal symmetries for the metric potentials are
finite, bounded and do not possess any singularity throughout the star and also at boundary, i.e for $0\le r \leq R_b$.
It is worthwhile to mention here that the value of Bag constant has been fixed to be $B=0.002$, which less than the predicted value of Bag constant $0.98<B<1.52$ in the units of $B_0=60 MeV/fm^3$ \cite{MIT4}, but otherwise we might not get some realistic results for analyzing compact stars in this study. Perhaps, this is due to the use of solutions through conformal symmetries.
 \item
The evolution of energy density and pressure functions is exhibited in Fig. \textbf{1}. Both energy density and pressure profiles behave realistically, except an unavoidable central singularity.
The $\rho$ is seen positive, decreasing to a minimum value at $r=R_{b}$ while pressure approaches to zero near the boundary of the star. The analysis is done with $R_{b}=9.36021,~9.30119,~9.20909,~9.14507,~9.12904$. The details are shown in table \textbf{I}. The derivatives of energy density and pressure functions with respect to radial coordinate are observed negative for the current study (Fig. \textbf{2}). The negativity in gradients shows that our obtained solutions are physically acceptable.
  \item We have discussed five kinds of energy bounds $TEC$, $WEC$, $NEC$, $SEC$,$DEC$ in this work (though one of them is abandoned now and only four are now being discussed in the literature). It is evident from Figs. \textbf{1, 3, 4} that all these energy conditions are satisfied in our case. Moreover for our curiosity, we have also investigated $TEC$ which turns out to be constant, i.e. $\frac{B}{2 \pi }>0$, which should turn out to be negative but not the case with neutron stars \cite{EC,EC1}.
\item
We have also discussed the equilibrium conditions through TOV equation for the current charged stellar structure study in the background of Bardeen geometry admitting conformal motions.
The interesting behavior of the hydrostatic, gravitational and electric forces can be seen balanced, which is shown in the right plot of Fig. \textbf{4}.
The non-dashed and the dashed curves above the axis represent hydrostatic and electric forces while the small dashed curves below the axis are to represent the gravitational force.
The balancing trend of all the above forces, i.e., ${F}_{\mathrm{h}}$, ${F}_{\mathrm{g}}$ and ${F}_{\mathrm{e}}$ indicates that our obtained electrically charge and conformal motion solutions for the compact stars are stable and physically acceptable.
\item The stability analysis of a compact star study is also an important feature. In this study, we obtain $v^2=1/3$ which is in a good agreement with the requirement to justify the causality condition. Moreover, the behavior of Adiabatic index $\gamma_{r}$ is shown in figure \textbf{5}. It is clearly shown that all the curves involved remain in a zone $>4/3$. Thus the results with $\gamma_{r}>4/3$ indicate that solutions with conformal motion with Bardeen model as an exterior geometry of star support stability criteria through Adiabatic index.
\item We have also provided a little review of the case with Reissner-Nordstrom spacetime as an exterior geometry for the matching condition. This enables us to provide some comparison with our study.
As a first step we keep the values of parameters $R_b,~ B$ and $C$ same as in the Bardeen model case, so that we may get an idea what happens to the fourth parameter $M$. The results are tabulated in Table \textbf{II}. It is interesting to notice that the values of radii remain same as in the Bardeen case, however, decrease in the values of mass is evident in Table \textbf{II} as compared to those in the case of Bardeen model (see Table \textbf{I}). Moreover, the variation corresponding to given radii is so small that we have to increase the number of decimal place accuracy to see a difference. As a second step, when we try to recover the same mass in the case of Bardeen model, we do not get realistic results. For example, for $(M,R_b)=(2.75064407,9.12904)$ we obtain both $B$ and $C$ negative $(B=-3.0429,~C=-253.594)$, providing a negative profile for energy density. For both Bardeen and RN models, the masses obey the Andreasson's limit $\sqrt{M} \leq \frac{\sqrt{R_b}}{3}+\sqrt{\frac{R_b}{9}+\frac{q^2}{3R_b}}$ requirement for a charged star \cite{Anderson}, (kindly see Tables \textbf{I} and \textbf{II}).

\item Finally, we have included some comparison with realistic structure model PSR J $1614-2230$ \cite{a1}. The M-R diagram in the light of present study for PSR J $1614-2230$ is shown in Fig. \textbf{6}. It is important to notice that realistic range of mass of PSR J $1614-2230$ depends upon the value of constant $C$. The higher values do not favor the mass of PSR J $1614-2230$ as in observed range. However, the smaller values (kindly see brown and red colored curves in Fig. \textbf{6}) exactly correspond to the masses in favorable domain. In fact, it is interesting to notice that the effects of electric charge with Bardeen geometry actually may improve the net mass of the structure. Thus more massive stellar structures can be theorized in the context of this study with conformal motion as compared to previous work \cite{a2}.
\end{itemize}

In nutshell, conformal symmetries are quite helpful in developing mathematical background for some physical solution in compact star study, with a drawback i.e., a singularity at the center. The calculated results using Bardeen geometry as an exterior spacetime are well-behaved in nature except for central singularity. In particular, the results show that Bardeen model case provides more massive stellar objects as compared to usual RN case. For example, stellar structure with masses greater than $2.93 M_{\odot}$ are obtained as compared to usual RN case where it was  $2.86 M_{\odot}$ (for details and comparison kindly see \cite{MAK1}). However, the integration constant $C$ has also an important role
(for details kindly see \textbf{II} and \cite{MAK1})). In fact, the current study supports the existence of realistic massive structures like PSR J $1614-2230$.



\section*{References}
\begin{thebibliography}{36}

\bibitem{MIT1} E. Witten, Phys. Rev. D \textbf{30} 272 (1984)

\bibitem{MIT0} G. Baym and S. A. Chin, Phys. Lett. \textbf{B62} 241 (1976).

\bibitem{MIT01} K. F. Liu and C. W. Wong, Phys. Lett. \textbf{B113} 1 (1982).



\bibitem{MIT2} E. Farhi and R. L. Jaffe, Phys. Rev. D \textbf{30} 2379 (1984)

\bibitem{MIT3} M. Malheiro, M. Fiolhais and A.R. Taurines, Journal of Physics G: Nuclear and Particle Physics \textbf{29} 1045 (2003)

\bibitem{3400} P. H. R. S. Moraes, J. D. V. Arbanil, and M. Malheiro: JCAP 2016.06 (2016) 005.


\bibitem{TOV1} J. R. Oppenheimer and G.M. Volkoff: Phys. Rev. 55, 374 (1939).
\bibitem{TOV2} J. Ponce de Leon: Gen. Relativ. Gravit. 25, 1123 (1993).

\bibitem{TOV3} J. Kumar, S. K. Maurya, A. K. Prasadand A. Banerjee: JCAP 2019(11), 005 (2019).


\bibitem{Ast2} A. V. Astashenok, S. Capozziello and S. D. Odintsov, J. Cosmol. Astropart. Phys. \textbf{2015}, 001 (2015).

\bibitem{Momeni} D. Momeni and R. Myrzakulov, Int. J. Geom. Methods Mod. Phys. 12, 1550014 (2015).

\bibitem{Ast3} A. V. Astashenok, S. Capozziello and S. D. Odintsov, Phys. Lett. B \textbf{742}, 160 (2015).

\bibitem{H14} L. Herrera and A. Di Prisco, Int. J. Mod. Phys. \textbf{D27}, 1750176 (2018)

\bibitem{H10} L. Herrera, J. Jimenez, L. Leal, J. Ponce de Leon, M. Esculpi and V. Galina, J. Math. Phys. \textbf{25}, 274(1984).
\bibitem{H11} L. Herrera and J. Ponce de Leon, J. Math. Phys. \textbf{26}, 778 (1985)
\bibitem{H12} L. Herrera and J. Ponce de Leon, J. Math. Phys. \textbf{26}, 2018 (1985)
\bibitem{H13} L. Herrera and J. Ponce de Leon, J. Math. Phys. \textbf{26}, 2302(1985)

\bibitem{Jamil}  F. Rahaman et al, Astrophys Space Sci \textbf{330} 249 (2010)

\bibitem{Rahman12}A. Das, F. Rahaman,  B. K. Guha, et al., Eur. Phys. J. \textbf{C76} 654 (2016)

\bibitem{Aktas} C. Aktas and I. Yilmaz, Gen Relativ Gravit \textbf{39}, 849, 62 (2007).

\bibitem{Jamil1} F. Rahaman et al. Astrophys Space Sci  \textbf{325}, 137 (2010)

\bibitem{MAK1} M. K. Mak and T. Harko, Int. J. Mod. Phys. D \textbf{13}, 149 (2004)

\bibitem{Rahman123} A. Das, F. Rahaman, B.K. Guha and S. Ray, Astrophys Space Sci \textbf{358} 36 (2015)

\bibitem{AbbasShahzad} G. Abbas and  M.R. Shahzad, Astrophys Space Sci \textbf{363} 251 (2018)

\bibitem{512} A. Banerjee, S. Banerjee, S. Hansraj, et al., Eur. Phys. J. Plus \textbf{132}, 150 (2017).

\bibitem{Esculpi} M. Esculpi and E. Aloma, Eur. Phys. J. C \textbf{67}, 521,32 (2010)

\bibitem{SW} M. Sharif and A. Waseem, Astrophys Space Sci  \textbf{364}, 189 (2019)

\bibitem{SW1} M. Sharif and H. Ismat Fatima, Int. J. Mod. Phys. \textbf{D25}, 1650083 (2016)

\bibitem{Bardeen} J. M. Bardeen, \textit{Non-singular general-relativistic gravitational collapse}, Proceedings
of GR-5, Tiflis, Georgia, U.S.S.R. page \textbf{174} (1968)

%
%
%
%
%
%
%
%
%
%
%
%
%
%
%
%
%
%

%
%
%
%



\bibitem{Moreno} C. Moreno and O. Sarbach, Phys. Rev. D \textbf{67}, 024028 (2003)

\bibitem{Zhou} S. Zhou, J. Chen, and Y. Wang, Int. Jour. Mod. Phys. D \textbf{21} 1250077 (2012)

\bibitem{MIT4} N. Stergioulas, Living Rev. Relativ. \textbf{6} 3 (2003)

\bibitem{MaK} M. K. Mak, and T. Harko, Int. J. Mod. Phys. D \textbf{13} 149 (2004).

\bibitem{Garcia} E. Ayon-Beato, A. Garcia: Phys. Lett. B493 (2000) 149.

\bibitem{Anderson} H.  Andreasson, Commun. Math. Phys. 288 (2009) 715.

\bibitem{Fernando1} S. Fernando, J. Correa: Phys. Rev. D 86, 64039 (2012)

\bibitem{Fernando2} S. Fernando: Int. J. Mod. Phys. D 26(2017)1750071.

\bibitem{Flachi} A. Flachi, J. Lemos: Phys. Rev. D 87, 024034 (2013)

\bibitem{Ulhoa} S. C. Ulhoa: Braz. Jour. Phys. 44, 380 (2014)

 \bibitem{Nordstrom} G. Nordstrom: \textit{On the Energy of the Gravitational Field in Einstein’s Theory}, Verhandl. Koninkl. Ned. Akad. Wetenschap., Afdel. Natuurk., 26 (1918) 1201.

%
%
\bibitem{Hil} W. Hillebrandt and K. O. Steinmetz, Astron. Astrophys. \textbf{53}, 283 (1976).

\bibitem{EC2} S. W. Hawking and G. F. R. Ellis, The Large Scale Structure of Space time (Cambridge University Press, Cambridge,
1975).
\bibitem{EC3} M. Visser, Science 276, 88 (1997).

\bibitem{EC4} S. Nojiri and S. D. Odintsov, Phys. Lett. B 571, 1 (2003).
\bibitem{EC5} J. Santos, J. S. Alcaniz, M. J. Reboucas and F. C. Carvalho, Phys. Rev. D 76, 083513 (2007).


\bibitem{EC6} O. Bertolami and M. C. Sequeira, Phys. Rev. D 79, 104010 (2009).
\bibitem{EC7} N. M. Garcia, T. Harko, F. S. N. Lobo and J. P. Mimoso, J. Phys. Conf. Ser. 314, 012060 (2011).

\bibitem{EC8} L. Balart and E. C. Vagenas, Phys. Lett. B 730, 14 (2014).
\bibitem{EC9} K. Atazadeh and F. Darabi, Gen. Relativ. Gravity 46, 1664 (2014).
\bibitem{EC10} S. Capozziello, S. Nojiri and S. D. Odintsov, Phys. Lett. B 781, 99 (2018).
\bibitem{EC11} P. Beltracchi and P. Gondolo, Physical Review D 99, 044037 (2019).

\bibitem{EC} Ya. B. Zeldovich and I.D. Novikov: Stars and Relativity (Relativistic Astrophysics, Vol. 1),
(University of Chicago Press, Chicago, 1971), see esp. p. 197.

\bibitem{EC1} M. Visser and C. Barcelo: Cosmo-99, 98, (2000).

\bibitem{a1} P. Demorest, T. Pennucci, S. Ransom, M. Roberts, J. Hessels, Nature 467, 1081 (2010).

\bibitem{a2} P. M. Takisa, S. D. Maharaj and L. L. Leeuw, Eur. Phys. J. C \textbf{79}, 8 (2019).


\end{thebibliography}
\end{document}